\documentclass[%
 reprint,
 amsmath,amssymb,
 aps,
]{revtex4-1}

\usepackage{graphicx,float}
\usepackage{dcolumn}
\usepackage{bm}
\usepackage[colorlinks=true,linkcolor=blue,citecolor=red,urlcolor=black,bookmarks=true,pdfstartview=FitB]{hyperref}

\begin{document}

\preprint{APS/123-QED}

\title{Coherent blue emission induced by a combination of diode and femtosecond lasers}

\author{Jesus P. Lopez}
\author{Marcio H. G. de Miranda}%
\author{Sandra S. Vianna}
\email{vianna@ufpe.br}
\affiliation{%
Departamento de F\'{\i}sica,
Universidade Federal de Pernambuco,
50670-901 Recife, PE - Brazil
}%

\author{Marco P. M. de Souza}
\affiliation{
 Departamento de F\'{\i}sica, Universidade Federal de Rond\^onia,
 76900-726 Ji-Paran\'a, RO - Brazil
}%

\date{\today}

\begin{abstract}
We report the investigation of a collimated blue light generated in rubidium vapor due to the combined action of an ultrashort pulse train and a cw diode laser. Each step of the two-photon transition 5S - 5P$_{3/2}$ - 5D is excited by  one of the lasers, and the induced coherence between the 5S and 6P$_{3/2}$ states is responsible for generating the blue beam. Measurements of the excitation spectrum reveal the frequency comb structure, indicating that each individual mode is responsible for inducing a nonlinear process. The strong signal dependency on the atomic density is characterized by a sharp growth and rapid saturation. 
\end{abstract}

\pacs{Valid PACS appear here}
\maketitle


Nonlinear interactions of light and atoms can be enhanced dramatically through the generation of quantum coherence among atomic states. In particular, atomic coherence effects have been explored in four-wave mixing to produce efficiently frequency up-conversion using either low power continuous wave (cw) lasers~\cite{Zib2002, Mei2006, Brekke2013} or pulsed lasers~\cite{Efthimiopoulos1996, Ariunbold2011}. Previous investigations demonstrated the high temporal coherence of the collimated blue light generated in Rb vapor~\cite{Mei2006} and the ability to transfer orbital angular momentum between the pump and generated beams ~\cite{Walker2012, Akulshin2015, Akulshin2016}. The interest in these investigations includes quantum information processing and memory~\cite{Cai2015}, photon correlation effects~\cite{McCormick2007, Willis2011} and tunable coherent sources~\cite{Aku2012}.

\begin{figure}[ht]
  \centering
  \includegraphics[width=7.5cm]{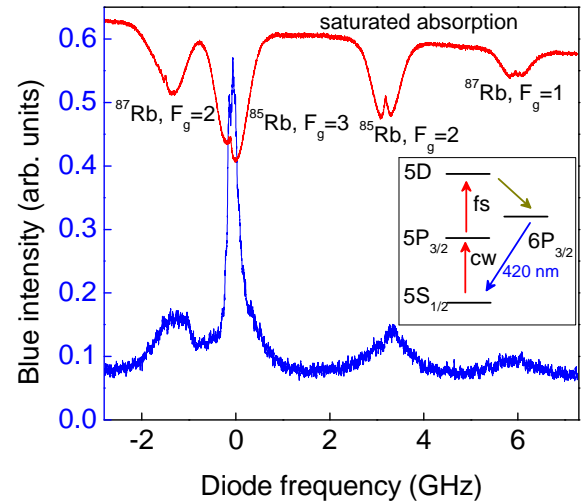}
  \caption {Excitation spectrum of the blue light emission as a function of the diode frequency for the D$_{2}$ Doppler lines at $T = 85$ $^o$C and $I_{cw} = 1.9$ W/cm$^2$. This curve is the average of two scans. The saturated absorption signal (upper curve) is detected simultaneously. The inset shows the relevant energy levels of Rb for the PFWM process.}
  \label{fig1}.
  \vspace{-0.2cm} 
  \end{figure}

In this work, we investigate the effects of the two-photon combined interaction of a cw laser and a mode-locked (fs) laser in an Rb vapor for the generation of coherent blue light. The two copropagating beams, at 780 nm (cw) and 776 nm (fs), drive each step of the two-photon transition 5S $_{1/2}$ $\rightarrow$ 5P$_{3/2}$ $\rightarrow$ 5D, respectively (see inset of Fig. \ref{fig1}). The induced coherence among 5S $_{1/2}$ $\rightarrow$ 5P$_{3/2}$ $\rightarrow$ 5D $\rightarrow$ 6P$_{3/2}$ transitions produces, by parametric four-wave mixing (PFWM), a coherent beam at 420 nm~\cite{Zib2002, Brekke2013, Akulshin2015}. Under the  coherent accumulation condition, where the atomic relaxation times are greater than the fs laser repetition period, we show that each individual mode of the frequency comb contribute to the nonlinear signal. The signature of this behavior is the observation of the frequency comb structure in the excitation spectrum of the coherent light, indicating that each mode is responsible for the generation of a blue beam with a frequency determined by the parametric process.  

The role of each mode of the frequency comb in the two-photon transition of Rb vapor has been investigated using similar experimental schemes,
with diode (cw) and mode-locked (fs) beams in a co- and contra-propaganting configurations, but detecting the fluorescence at 90$^{0}$~\cite{Mor2012, Fil2015}. In both cases, the detected signal has its origin in a spontaneous emission process and reflects the population of the excited state 5D. In the PFWM process investigated here, the nonlinear signal is determined by two-photon coherence between the 5S and 5D states, and by the amplified spontaneous emission at 5.2 $\mu$m~\cite{Aku2009}. In this case, the generated blue light reflects not only the characteristics of the atomic system, but also carries the phase information related with the excitation beams. 
 
In the experiment a diode laser, stabilized in temperature and with a linewidth of about $1$~MHz, is used to excite the $5S_{1/2} \rightarrow 5P_{3/2}$ transition. A train of pulses generated by a mode-locked Ti:sapphire laser (MIRA, Coherent) can excite both $5S_{1/2} \rightarrow 5P_{3/2}$ and $5P_{3/2} \rightarrow 5D$ transitions. The two copropagated beams, with parallel linear polarizations, are focused in a 5-cm long sealed Rb vapor cell. The vapor cell is heated up to 
$\approx 100~^0$C and contains both $^{85}$Rb and $^{87}$Rb isotopes in their natural abundances. We measure the dependence on the Rb density of the collimated blue light generated in the forward direction for different diode laser intensities. The measurements were performed as a function of the diode frequency allowing us to realize a velocity-selective spectroscopy~\cite{Aumiler2005, Polo2011}.

The Ti:sapphire laser with a repetition rate of $f_{R} \approx 76$ MHz produces $100$~fs pulses and $500$~mW of average power. The fs laser intensity was kept fixed with a mode intensity of order of 1.0 mW/cm$^{2}$ in the cell entrance. The diode laser can sweep over $10$~GHz by tuning its injection current and a saturated absorption setup is used to calibrate its frequency. The diameter of the two beams is almost constant inside the cell and it is about $400$ $\mu$m for the fs beam and $260$ $\mu$m for the diode laser. The blue beam generated at 420 nm is collected in the forward direction, satisfying the phase-matching condition in the parametric four-wave mixing process. Bandpass filters and a spectrometer, placed about 1.0 m from the cell, are used to cut the light around 780 nm. The signal is detected with a photomultiplier tube and recorded on a digital oscilloscope. 

\begin{figure}[ht]
  \centering
  \includegraphics[width=7.5cm]{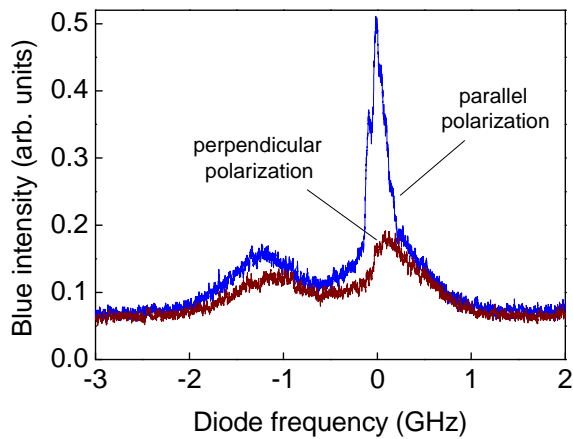}
  \caption {Blue light intensity as a function of the diode frequency for two analyzer polarizations:  parallel (blue curve) and perpendicular (red curve) to the incident beams polarization, at $T = 85^o$C and $I_{cw} = 1.9$ W/cm$^2$. The peak in the parallel polarization verifies the PFWM process.}
  \label{fig2}.
  \vspace{-0.2cm} 
  \end{figure} 

Figure \ref{fig1} shows the blue signal intensity (lower curve), for a fixed $f_{R}$, as the diode frequency is scanned over the four Doppler-broadened $D_{2}$ lines of the $^{85}$Rb and $^{87}$Rb. The origin of the horizontal scale was arbitrarily chosen at the $5S_{1/2}, F_{g} = 3 \rightarrow 5P_{3/2}, F_{g} = 4$ transition of the $^{85}$Rb. The spectrum consists of one sharp and intense peak over a weak broad peak together with other three weak broad peaks, all over a flat background. The intense peak is due to the PFWM process when the two-photon transition is excited by both lasers: the diode laser and the fs laser. The weak broad peaks are the blue fluorescence induced by both lasers; while the flat background is due only to the excitation by the fs laser~\cite{Ban2013}. The spectrum was obtained for cell temperature of 85~$^{o}$C and diode intensity of $I_{cw} = 1.9$ W/cm$^2$. The saturated absorption curve (upper curve) is used to calibrate the diode frequency. In Fig. \ref{fig2}  we display the excitation spectra when the generated signal is detected after an analyzer. We clearly see that while the fluorescence is not polarized, the intense peak due the PFWM process practically disappears when the polarization of the incident beams and the analyzer are perpendicular, indicating that this signal is generated with the same polarization of the incident beams~\cite{Kienlen2013}. This polarization behavior is also a demonstration that the blue emission has its origin in the nonlinear process.

\begin{figure}[ht]
  \centering
  \includegraphics[width=7.5cm]{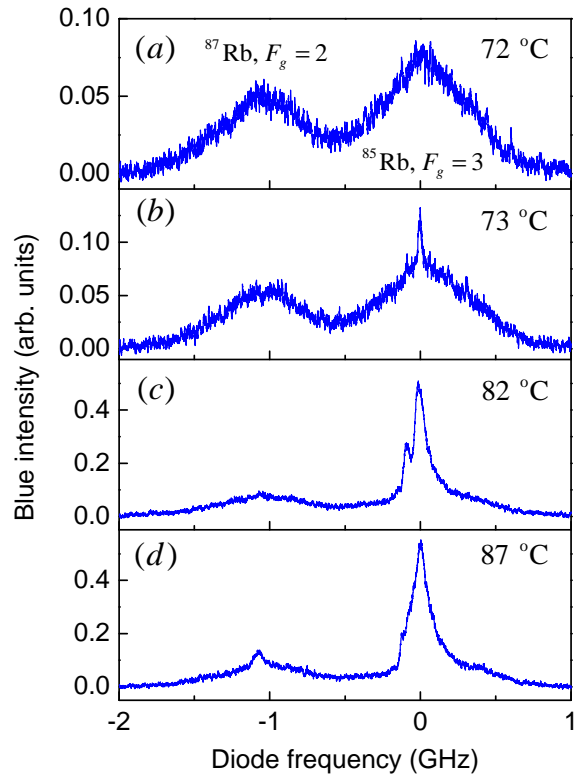}
  \caption {Excitation spectrum of the blue light emission as a function of the diode frequency for different temperatures. $I_{cw} = 1.9$ W/cm$^2$ and $I_{fs} = 1.0$ mW/cm$^2$ (each mode). The flat background was removed.}
  \label{fig3}.
  \vspace{-0.2cm} 
  \end{figure}

The blue signal for different Rb vapor temperatures is shown in Fig. \ref{fig3} for the Doppler lines $F_{g}=3$ of $^{85}$Rb and $F_{g}=2$ of $^{87}$Rb when the diode laser intensity is fixed at $I_{cw} = 1.9$ W/cm$^2$. For comparison, the flat background has been removed.  We note from Figs. \ref{fig3}(a) and \ref{fig3}(b) that, for $^{85}$Rb, the threshold of the PFWM process occurs between 72 and 73 $^o$C. The critical dependence of the PFWM signal on the atomic density is revealed when only one  mode of the frequency comb contributes to the signal. This mode is close to resonance with the cyclic transition for the atomic velocity group that has the highest density. 

By increasing temperature other velocity groups reach the threshold of the atomic density, making possible more modes to contribute to the PFWM signal. In Fig. \ref{fig3}(c) we can distinguish the signal due to two modes of the frequency comb. At higher temperatures, the threshold for the $^{87}$Rb density is reached, as showed by the thin peak near -1 GHz in Fig. \ref{fig3}(d). Under these conditions,  we note in Fig. \ref{fig3}(d) that the mode structure for the $^{85}$Rb appear blurred due to saturation and absorption effects of the most intense peaks, along with a rapid scanning of the diode laser, resulting in a broad line. This line shape is similar to that is obtained when two diode lasers are tuned to the maximum of the blue light power~\cite{Aku2009}. This critical atomic density dependence explains why is hard to observe the comb structure in the PFWM signal, when compared with the fluorescence signal~\cite{Fil2015}.

\begin{figure}[ht]
  \centering
  \includegraphics[width=6.5cm]{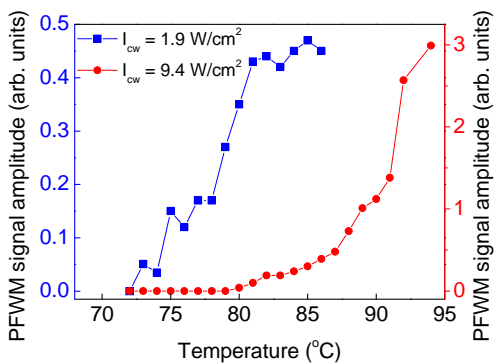}
  \caption {Amplitude of the PFWM signal in the Doppler line $F_{g}=3$ of $^{85}$Rb as a function of the temperature, for two diode laser intensities.}
  \label{fig4}.
  \vspace{-0.2cm} 
  \end{figure}

The PFWM process presents different atomic density thresholds for different laser intensities. In Fig. \ref{fig4} we present the PFWM signal amplitude as a function of the cell temperature for two diode intensities: $I_{cw} = 1.9$ W/cm$^2$ (blue squares) and $I_{cw} = 9.4$ W/cm$^2$ (red circles). To get the amplitude of the signal, we fit each spectrum with different Gaussian curves, in order to separate the fluorescence of each peak related with the PFWM process. For $I_{cw} = 1.9$ W/cm$^2$, we see a threshold at $\approx 72 ^o$C and  a saturation at $\approx 81 ^o$C, while for $I_{cw} = 9.4$ W/cm$^2$ these temperatures change to $\approx 80 ^o$C and $\approx 95 ^o$C, respectively. This threshold atomic density behavior is very similar to what has been observed using only cw diode lasers ~\cite{Aku2009}. However, the increase of the atomic density threshold while the diode intensity increases, observed in the experiment described here, indicates a competition between an optical pumping effect and the two-photon absorption, probably due to the low intensity of the frequency comb, $I_{fs} = 1.0$ mW/cm$^2$ (each mode). We also understand that the low intensity of each mode of the frequency comb is responsible for the nonobservation of a collimated blue emission far detuning from the one-photon resonance at hight temperature~\cite{Mei2006, Aku2009}.

\begin{figure}[ht]
  \centering
  \includegraphics[width=7.5cm]{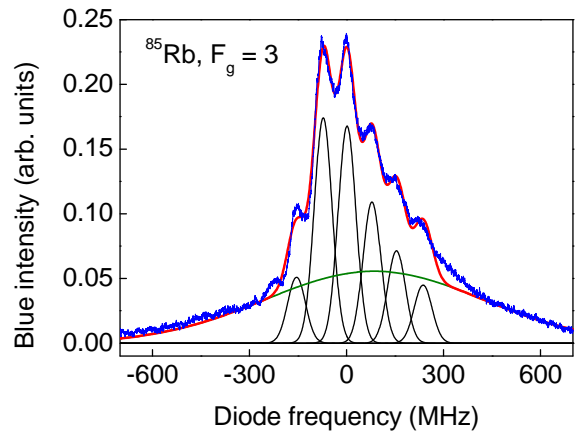}
  \caption {Blue light intensity (blue line) and fit curves (red, green and black lines) as a function of the diode laser frequency, at $T = 87$ $^o$C. The red line is the sum of the fluorescence contribution (green curve) and the contribution of the six modes to the PFWM signal (black lines). The blue curve is the average of 10 measurements.}
  \label{fig5}.
  \vspace{-0.2cm} 
  \end{figure}

	The fit discussed above is illustrated in Fig. \ref{fig5}, where we have the experimental blue signal (blue line) obtained at $T = 87 ^o$C. To visualize the several modes of the frequency comb printed in the Doppler profile, we perform a slow scan of the diode laser with an average of 10 measurements. 
The green line corresponds to a Gaussian curve that fit the weak broad peak of the fluorescence, while the black curves correspond to Gaussian curves, all with same linewidth but dislocated in frequency, that fit the PFWM signal.  The red line is the sum of all Gaussian curves and, as we can see, describes well our experimental blue signal. The frequency difference between two adjacent black peaks is $78\pm 4$ MHz, close to the $f_R$ of the laser, and its linewidths are $\approx$ 55 MHz. This result indicates that the peak structure observed is a clear contribution of each mode to the induced coherence in the vapor, which is responsible for generating the coherent blue light. Each mode excites a different group of atoms, making the process selective in atomic velocity.

  	In conclusion, we have investigated the coherent blue light generated in atomic Rb vapor using a parametric four-wave mixing process due to the combined action of a cw laser and a train of ultrashort pulses. The frequency comb, that drives the upper transition, behaves like a set of cw diode lasers with different frequencies, and the atomic system interacts resonantly only with a few of them, resulting in an excitation selective of different atomic velocity groups. For each mode, the atomic density dependence is characterized by a sharp growth and rapid saturation. Remarkably, only one scan of the diode laser allows us to generate several blue beams, each one from a different atomic velocity group with its blue frequency determined by the parametric four-wave mixing process. Studies of the spectral and temporal characteristics of the blue emission are in development. Another feature to be explored is the fixed phase relation between the modes of the frequency comb and the possibility to transfer this phase relation to the generate beams.	

\vspace{1.0cm} \noindent \textbf{\large{Funding}}. Conselho Nacional de Desenvolvimento Cient\'{\i}fico e Tecnol\'ogico (CNPq), Coordena\c{c}\~ao de Aperfei\c{c}oamento de Pessoal de N\'{\i}vel Superior (CAPES), Funda\c{c}\~ao de Amparo \`{a} Ci\^encia e Tecnologia de Pernambuco (FACEPE).


\end{document}